# Superconducting and thermoelectric properties of new layered Superconductor $Bi_4O_4S_3$

S. G. Tan[a], L. J. Li[a], Y. Liu[a], P. Tong[a], B. C. Zhao[a], W. J. Lu[a], Y. P. Sun[a,b,*]

[a] Key Laboratory of Materials Physics, Institute of Solid State Physics, Chinese Academy of Sciences, Hefei 230031, People's Republic of China

[b] High Magnetic Field Laboratory, Chinese Academy of Sciences, Hefei 230031, People's Republic of China

**Abstract**

Polycrystalline sample of the new layered superconductor $Bi_4O_4S_3$ is successfully synthesized by solid-state reaction method by using Bi, S and $Bi_2O_3$ powders with one step reaction. The superconducting transition temperature ($T_c^{onset}$=4.5 K), the zero resistance transition temperature ($T_{c0}$=4.07 K) and the diamagnetic transition temperature (4.02 K at $H$=10 Oe) were confirmed by electrical transport and magnetic measurements. Also, our results indicate a typical type II-superconductor behavior. In addition, a large thermoelectric effect was observed with a dimensionless thermoelectric figure of merit (ZT) of about 0.03 at 300K, indicating $Bi_4O_4S_3$ can be a potential thermoelectric material.

*Corresponding author*: Tel: +86-551-559-2757; Fax: +86-551-559-1434.
**E-mail:** ypsun@issp.ac.cn

To study the layered materials is one of the strategies for exploring new superconductors. The discoveries of superconductivity with high superconducting transition temperatures in cuprates[1-4] and oxypnictides[5-8] are good examples. Very recently, Mizuguchi et al., reported the superconductivity in layered Bismuth-oxysulfide $Bi_4O_4S_3$[9] with $T_{c0}$=4.5 K. Just ten days later, S. Li et al., reported the measurements of resistivity, Hall effect and magnetization of $Bi_4O_4S_3$.[10] Their results indicate an exotic multi-band behavior and show the feature of superconducting pairing in one dimensional chains. Structurally, $Bi_4O_4S_3$ is composed of stacking of $Bi_4O_4(SO_4)_{1-x}$ and $Bi_2S_4$ layers. The $BiS_2$ layer is a basic unit as the Cu-O layer in Cu-based superconductors[1-4] and the Fe-An (An=P, As, Se, Te) layer in oxypnictides.[5-8] The $Bi_4O_4S_3$ superconductor as a new member in the layered superconducting family may stimulate further studies on the materials with $BiS_2$ layer.

The reported preparations for $Bi_4O_4S_3$ superconductor contains two steps.[9-10] Firstly, the $Bi_2S_3$ polycrystalline powder should be prepared using solid-state reaction by Bi powder and S grain. And then, the obtained $Bi_2S_3$ powder, $Bi_2O_3$ powder and S grains were used to prepare $Bi_4O_4S_3$ polycrystalline by solid-state reaction. In this letter, we report a one-step solid-state reaction method for the superconducting $Bi_4O_4S_3$. Our measurement confirms $Bi_4O_4S_3$ is a typical type II-superconductor by magnetization measurement. In addition, we report a large thermoelectric effect in this compound, which may excite further studies of thermoelectric effect in those compounds with $BiS_2$ layers for the purpose of potential applications.

The Bi, S and $Bi_2O_3$ powders in stoichiometric ratio were fully mixed and ground, and then pressed into pellets. The pellets were sealed in an evacuated silica tube. The tube was heated for 10 hours at 510 °C. The obtained pellets were ground again and the above treatment was repeated. Finally, polycrystalline $Bi_4O_4S_3$ were obtained. The room-temperature crystal structure and lattice constants were determined by powder X-ray diffraction (XRD) (Philips X'pert PRO) with Cu $K_\alpha$ radiation. The electronic and thermal transport measurements were performed in a Quantum Design physical property measurement system (PPMS), and the magnetization measurement was performed on a superconducting quantum interference device (SQUID) system.

Figure 1 shows the XRD pattern for the $Bi_4O_4S_3$ polycrystalline. The Bragg diffractions can be indexed using the tetragonal structure with the space group of $I4/mmm$. The lattice parameters are obtained by fitting the powder XRD pattern by using the Rietica software,[11] and the obtained lattice constants *a*, *b* and *c* are $a=b=0.3978$nm, $c=4.107$nm, respectively, which are in consistence with the reported results.[9]

Figure 2 shows the temperature dependence of the resistivity of the $Bi_4O_4S_3$ polycrystalline measured by a standard four-probe method under a zero magnetic field. As seen from Fig. 2, the $Bi_4O_4S_3$ has a metallic behavior in the normal state, and the magnitude of the resistivity is slightly less than that reported in Ref. 9 in the whole measurement temperature region. When the sample was cooled down to 5 K, the resistivity has an abrupt drop due to the occurrence of superconductivity. The inset of Fig. 2 shows the low temperature resistivity of $Bi_4O_4S_3$ at different applied magnetic field (*H*). The superconducting transition temperature ($T_c^{onset}$) and zero resistivity temperature ($T_{c0}$) of the sample are determined to be 4.5 K and 4.02 K at H=0T, respectively, and the superconducting transition width is determined to be 0.3 K according to 10–90% of the normal state resistivity, such a narrow width indicates the homogeneity of the sample. The $T_c^{onset}$ and $T_{c0}$ shift to lower temperatures with the increase of *H*.

The superconductivity was also proved by the magnetic measurements shown in Fig. 3. The diamagnetism in the low-temperature region further confirms the existence of superconductivity, and the steep transition in the *M* (*T*) curve indicates that the sample is rather homogeneous. The superconducting diamagnetic transition begins at 4.02 K, defined by the onset point of the zero-field-cooling (ZFC) and field-cooling (FC) curves. It is consistent with the $T_{c0}$ obtained from the resistance measurement. The smaller magnetization value for FC is likely due to the complicated magnetic flux pinning effects.[12] The value of $-4\pi\chi$ at 2 K is about 33%. The top inset of Fig. 3 shows the magnetization hysteresis loop of $Bi_4O_4S_3$ at *T*=2K. The shape of the *M* (*H*) curve indicates that $Bi_4O_4S_3$ is a typical type-II superconductor. The bottom inset of Fig. 3 shows the initial *M* (*H*) curves of the $Bi_4O_4S_3$ in the low-field region at 2 K, which allows us to estimate the lower critical field values ($H_{c1}$) at 2 K. At low fields, the *M* (*H*) isotherm is linear with *H*, as expected for a BCS type-II superconductor. The estimated $H_{c1}$ (2 K) value is about 7.4 Oe, defined by the point where the curve deviates from linearity (marked by the arrow in the inset of Fig.3).

The thermal transport measurement results which are containing the Seebeck coefficient (*S*), electrical resistivity *r*, thermal conductivity *k* are shown in Fig. 4. The

$k$ (T) shows a broad peak at around 25K. The origin is unknown at present. The $S$ (T) is negative in the whole measurement temperature region, indicating the major carriers are electron-like. It is in agreement with the Hall measurement results.[10] The $S$ (T) is linearly dependent with temperature when $T$ is less than 50K, above which the $S$ (T) is less temperature dependent. It results in a broad slope change at ~ 100K. In the similar temperature region, a broad minimum in the Hall coefficient $R_H$(T) was observed.[10] It indicates the temperature dependence of $S$ (T) is mainly governed by the evolution of charge carrier density with temperature. The efficiency of a thermoelectric material is determined by the dimensionless figure of merit, ZT (ZT=$S^2T/\rho\kappa$). Hence, a good thermoelectric material should have high $S$, low $\kappa$ and $\rho$. The dimensionless figure of merit (ZT) for $Bi_4O_4S_3$ is presented in Fig. 4(b). Due to low resistivity, and relatively small thermal conductivity, the ZT value of our sample at 300K reaches about 0.03. The ZT value at 300K is comparable to the Fe-based superconductors $LaFePO_{1-x}F_x$ and $LaFeAsO_{1-x}F_x$,[13] while larger than cuprates HgBaCaCuO (1223) and BiSrCaCuO (2212)[14] by a few magnitudes. It is interesting to modify the $S$ (T), $\kappa$ (T) and $\rho$ (T) to increase the ZT value so as to meet the requirements of practical applications, for instance, via chemical doping.

In summary, polycrystalline samples of the new layered superconductor $Bi_4O_4S_3$ are prepared by a one-step solid-state reaction method using Bi, S and $Bi_2O_3$ powders. The electronic and magnetic measurements show that the onset transition temperature and zero resistance transition temperature are $T_c^{onset}$=4.5 K and $T_{c0}$=4.02 K, respectively. Our $M$ ($H$) measurement indicates $Bi_4O_4S_3$ is a typical type II-superconductor. The large thermoelectric effect with ZT=0.03 at room temperature, which may open a new scope for pursuing new thermoelectric materials with large ZT values.

**Acknowledgments:**

This work was supported by the National Key Basic Research under contract No. 2011CBA00111, and the National Nature Science Foundation of China under contract Nos. 10804111, 10974205, 11104279, 11174293 and 51102240, and Director's Fund of Hefei Institutes of Physical Science, Chinese Academy of Sciences.

**References:**

[1] J. G. Bednorz and K. A. Muller, Z. Physik B **64,** 189 (1986).

[2] M. K. Wu, J. R. Ashburn, C. J. Torng, P. H. Hor, R. L. Meng, L. Gao, Z. J. Huang, Y. Q. Wang, and C. W. Chu, Phys. Rev. Lett. **58,** 908 (1987).

[3] H. Maeda, Y. Tanaka, M. Fukutomi and T. Asano, Jpn. J. Appl. Phys. **27,** L209 (1988).

[4] P. Lejay, P. de Rango, A. Sulpice, B. Giordanengo, R. Tournier, R. Retoux, S. Deslandes, C. Michel, M. Hervieu and B. Raveau, Revue Phys. Appl. **24,** 485 (1989).

[5] M. Rotter, M. Tegel, and D. Johrendt, Phys. Rev. Lett. **101,** 107006(4pp) (2008).

[6] J. H. Tapp, Z. J. Tang, B. Lv, K. Sasmal, B. Lorenz, P. C. W. Chu, and A. M. Guloy, Phys. Rev. B **78,** 060505(R) (4pp) (2008).

[7] X. Y. Zhu, F. Han, P. Cheng, G. Mu, B. Shen, L. Fang, and H. H. Wen, EPL **85**, 17011 (2009).

[8] X. H. Chen, T. Wu, G. Wu, R. H. Liu, H. Chen, D. F. Fang, Nature **453,** 761 (2008).

[9] Y. Mizuguchi, H. Fujihisa, Y. Gotoh, K. Suzuki, H. Usui, K. Kuroki, S. Demura, Y. Takano, H. Izawa, and O. Miura, arXiv:1207.3145.

[10] S. Li, H. Yang, J. Tao, X. X. Ding and H. H. Wen, arXiv:1207.4955v1.

[11] J. A. Wilson, F. J. DiSalvo, S. Mahajan, Adv. Phys. **24,** 117 (1975).

[12] D. E. Prober, M. R. Beasley and R. E. Schwall, Phys. Rev. B **15,** 5245 (1977).

[13] T. Okuda, W. Hirata, A. Takemori, S. Suzuki, S.Saijo, S. Miyasaka, and S. Tajima, J. Phys. Soc. Jpn. **80**, 044704 (2011).

[14] H. Bougrine, M. Ausloos, R. Cloots and M. Pekala, Proc. 17th International Conference on Thermoelectrics, ICT98, Nagoya, Japan (IEEE, Piscataway, 1998) p. 273-276.

**Figure captions:**

Fig. 1 Rietveld refinement results of the XRD pattern at room temperature for the $Bi_4O_4S_3$ sample. Solid crosses indicate the experimental data and the calculated data is the continuous line overlapping them. The lowest curve is the difference between the experimental and calculated patterns. The vertical bars indicate the Bragg reflection positions. The Miller indices are also marked. Inset shows the crystal structure of $Bi_4O_4S_3$.

Fig. 2 Temperature dependence of the resistivity $\rho$ (T) for $Bi_4O_4S_3$. The inset shows the low-temperature $\rho$ (T) curve at different magnetic fields.

Fig. 3 Temperature dependence of the magnetic susceptibility of $Bi_4O_4S_3$ at $H$=10 Oe. The top inset shows the magnetization hysteresis loop of $Bi_4O_4S_3$ at $T$=2 K. The right bottom inset shows the initial $M$ ($H$) isotherm at $T$=2 K, the red dash line shows the linear fitting in the low field range.

Fig. 4 Temperature dependence of the (a) thermal conductivity and resistivity, (b) Seebeck coefficient and ZT values for $Bi_4O_4S_3$.

Figure 1

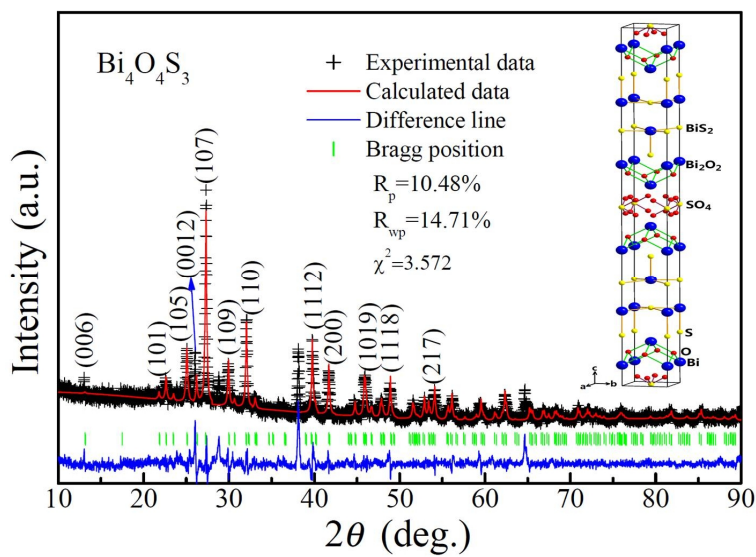

**Fig. 1** Rietveld refinement results of the XRD pattern at room temperature for the $Bi_4O_4S_3$ sample. Solid crosses indicate the experimental data and the calculated data is the continuous line overlapping them. The lowest curve is the difference between the experimental and calculated patterns. The vertical bars indicate the Bragg reflection positions. The Miller indices are also marked. Inset shows the crystal structure of $Bi_4O_4S_3$.

Figure 2

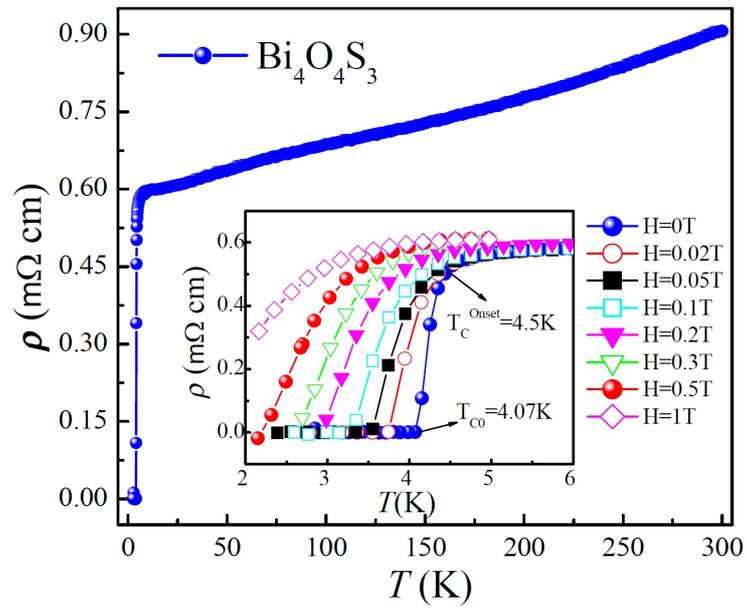

**Fig. 2** Temperature dependence of the resistivity $\rho$ (T) for $Bi_4O_4S_3$. The inset shows the low-temperature $\rho$ (T) curves at different applied magnetic fields.

Figure 3

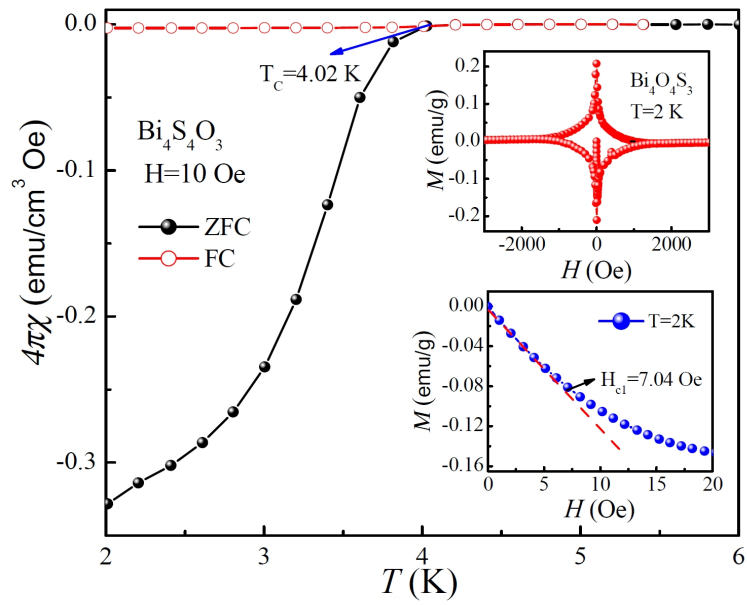

**Fig. 3** Temperature dependence of the magnetic susceptibility of $Bi_4O_4S_3$ at $H$=10 Oe. The top inset shows the magnetization hysteresis loop of $Bi_4O_4S_3$ at $T$=2 K. The right bottom inset shows the initial $M$ ($H$) isotherm at $T$=2 K, the red dash line shows the linear fitting in the low field range.

Figure 4

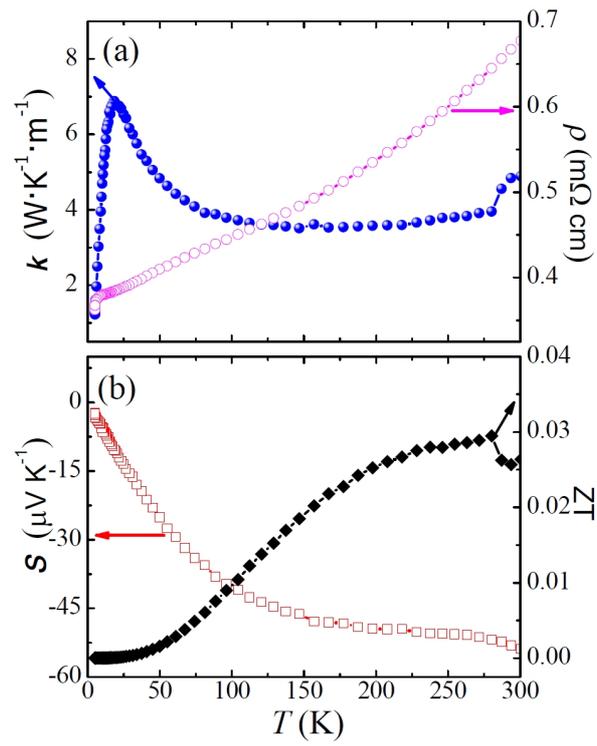

**Fig. 4** Temperature dependence of the (a) thermal conductivity and resistivity, (b) Seebeck coefficient and ZT values for $Bi_4O_4S_3$.